\documentclass[preprint]{revtex4}
\usepackage{epsfig}
\usepackage{amsfonts}
\usepackage{latexsym}
\usepackage{amsmath,amssymb}
\usepackage{mathrsfs}
\usepackage{hyperref}
\usepackage{setspace}
\usepackage{color}
\usepackage{bm}
\usepackage{slashed}
\usepackage{braket}
\begin{document}

\title{Harvesting quantum coherence  from axion dark matter}

\author{Sugumi Kanno$^1$, Akira Matsumura$^1$, Jiro Soda$^{2}$}

\affiliation{$^{1}$  Department of Physics, Kyushu University, 744 Motooka, Nishi-ku, Fukuoka 819-0395, Japan}
\affiliation{$^{2}$  Department of Physics, Kobe University, Kobe 657-8501, Japan}

\begin{abstract}
Quantum coherence is one of the most striking features of quantum mechanics rooted in the superposition principle. Recently it has been demonstrated that it is possible to harvest the quantum coherence from a coherent scalar field. In order to explore a new method of detecting axion dark matter, we consider a point-like Unruh-DeWitt detector coupled to the axion field and quantify a coherent measure of the detector. We show that the detector can harvest the quantum coherence from the axion dark matter. To be more precise, we consider a two-level of electron system in an atom as the  detector. 
In this case, we obtain the coherence measure $C=2.2\times 10^{-6}\gamma\, \left(T/1{\rm s}\right)$ where $T$ and $\gamma$ are an observation time and the Lorentz factor. At the same time, the axion mass $m_a$ we can probe is determined by the energy gap of the detector.
\end{abstract}

\maketitle
\tableofcontents
\section{Introduction}

In cosmology, to reveal the nature of the dark matter is the most important issue. 
Recently, axions~\cite{Peccei:1977hh,Weinberg:1977ma,Wilczek:1977pj,Kim:1979if,Shifman:1979if,Dine:1981rt}
 and axion-like particles~\cite{Svrcek:2006yi,Arvanitaki:2009fg} with broad range of  masses
 have been intensively studied as candidates for dark matter~\cite{Preskill:1982cy,Abbott:1982af,Dine:1982ah,Marsh:2015xka}. 
It is known that the energy density of the dark matter near the Earth is $0.3$ GeV/cm$^3$~\cite{Gorbunov:2011}. This means that 
in the case of the very light axion mass, say, $10^{-24}$ eV/$c^2$, the occupation number of the axion becomes very large ($\sim 10^{100}$) and naively we regard the axion as a classical field. From the observations of dwarf galaxies and distant galaxies, there are constrains on the mass of the axions larger than $10^{-21}$eV/$c^2$ ~\cite{Safarzadeh:2019sre,deVega:2014wya,Castellano:2018ogp,Menci:2017nsr}.
The observations of steller mass black holes with spins exclude the mass range between $10^{-20}$ eV/$c^2$ and $2\times 10^{-11}$ eV/$c^2$~\cite{Stott:2018opm,Stott:2020gjj}.
However, from observations of the large scale structure of the universe, there are no constraints on the axion mass between $10^{-10}
$ eV/$c^2$ and $1$ keV/$c^2$ . For the axion mass larger than $1$ keV/$c^2$, it is known that the axion annihlates into photons.
Hence, in this paper, we focus on the axion mass between $10^{-10}$ eV/$c^2$ and $1$ keV/$c^2$. In particular, we consider whether quantum coherence of the axion in the mass range can be harvested by a detector. 

A peculiar feature of the axion field is its coherent oscillation. This oscillation has a frequency given by the axion mass. Taking advantage of this feature of axion that behaves as waves, there are some attempts to detect this coherent oscillation directly with gravitational wave interferometers~\cite{Aoki:2016kwl,Yoshida:2017cjl,Jung:2020aem}.
Interestingly, the methods of detecting the axion and gravitational waves are similar~\cite{Ikeda:2021mlv,Ito:2019wcb}.
It is shown that quantum entanglement is useful for detecting gravitons~\cite{Kanno:2021gpt}.
The Hanbury-Brown-Twiss interferometer which is developed in quantum information science can be used to probe gravitons of primordial gravitational waves~\cite{Kanno:2018cuk}.  
Hence, it would be intriguing to make use of the method developed in quantum information science for probing
the axion dark matter by detecting the quantum coherence of the axion. Although there is an extensive amount of research on harvesting quantum entanglement from quantum fields~\cite{Reznik:2002fz}, no-go theorem for the entanglement harvesting is also discussed in~\cite{Simidzija:2018nku,Matsumura:2021tlu}.
The more fundamental feature of quantum physics is the superposition principle, namely quantum coherence. Recently, the importance of the quantum coherence has been recognized~\cite{Streltsov:2016iow}
and a coherence measure is invented~\cite{Marvian:2016upl}.
Thus, the goal of this paper is to study whether we can probe the axion dark matter
by harvesting the quantum coherence from it.

In macroscopic systems, decoherence occurs rapidly through ordinary interactions, such as electromagnetic interactions. However, the dark matter has very weak interactions with the standard model particles. The gravity couples to any sources of energy and momenta and thus couples to the dark matter with the same strength of the gravitational constant. 
Hence,  macroscopic quantum superposition states of the dark matter may 
experience decoherence through gravitational interactions. However, gravitation is so weak that the decoherence may proceed very slowly. Thus, the dark matter may keep its quantum coherence for a long time~\cite{Allali:2021puy}. 

In this paper, we focus on light axion fields. Since the dark matter is non-relativistic, the dark matter particles have a large de Broglie wavelength. 
The ultra-light axions with the mass from $10^{-10}$ eV/$c^2$ to $1$ keV/$c^2$ correspond to the de Broglie wavelengths
from $10^{9}$cm to submillimeters.
Thus, unlike typical de Broglie wavelengths associated with well-known particles (electrons and  protons etc.), the de Broglie wavelength of the light dark matter can be macroscopically large. 
 
Recently, a way of faithfull extraction of quantum coherence has been investigated in~\cite{Kollas:2020pji}.
Then, some concrete models have been considered for harvesting the quantum coherence~\cite{Kollas:2021nqy} (see also \cite{Kollas:2020ghd}). 
 In this paper, we apply the method that uses a coherence measure developed in quantum information science to the axion dark matter and
 investigate whether we can harvest the quantum coherence from the axion dark matter. 
 
The organization of the paper is as follows:
In section 2, we summarize the basics about the axion dark matter.
In section 3, we review the quantum coherence and the coherence measure.
In section 4, we investigate whether a detector can harvest the quantum coherence from the axion dark matter. The last section is devoted to the conclusion. 

\section{Axion dark matter background}
\label{section2}

We consider axions that may account for some or all of the missing dark matter in the universe. In the local galaxy, we can neglect the cosmic expansion and discuss the axion field in the Minkowski space.
The axion is a pseudo-scalar with a shift symmetry $\phi \rightarrow \phi + {\rm constant}$. The mass of the axion is induced by the noperturbative effect. 
The model is described by the action
\begin{eqnarray}
S= \int d^4 x \left[-\frac{1}{2}\partial^\mu \phi \partial_\mu \phi 
-\frac{1}{2}m_a^2  \phi^2 \right]\,,
\end{eqnarray}
where $m_a$ is the mass of the axion.
The axion $\phi$ is coherently oscillating scalar field with the frequency close to the axion's mass $\omega\approx m_a$ expressed by
\begin{eqnarray}
\phi = A \cos m_a t  \ , 
\label{axion}
\end{eqnarray}
where $A$ is the amplitude of the axion oscillation. 
Note that the energy density of the axion is given by 
\begin{eqnarray}
\rho_a=\frac{1}{2}\dot{\phi}^2+\frac{m_a^2}{2}\,\phi^2=\frac{1}{2}A^2 m_a^2 \ .
\end{eqnarray} 
where a dot denotes derivative with respect to $t$. If the dark matter consists of the axion, the energy density $\rho_a$ is regarded as $\rho_{\rm DM}$ and the amplitude of the axion can be estimated as
\begin{eqnarray}
A = \frac{\sqrt{2\rho_{\rm DM}}}{m_a} 
= 2\times10^{-6} \sqrt{\frac{\rho_{\rm DM}}{0.3\,{\rm GeV}/{\rm cm}^3}} \left(\frac{10^{-6}\,{\rm eV}}{m_a} \right)
             {\rm GeV} \ , 
\label{A}
\end{eqnarray}
where we used the energy density of the dark matter near the Earth 
$\rho_{\rm DM} = 0.3\,{\rm  GeV}/{\rm cm}^3$~\cite{Gorbunov:2011} and considered the light axion mass of $10^{-6}$ eV for reference. Dark matter is pressureless by definition and the existance of the pressureless matter is confirmed by the CMB observation~\cite{Planck:2015fie}. From Eq.~(\ref{axion}), the pressure of axion is given by
\begin{eqnarray}
p_a=\frac{1}{2}\dot{\phi}^2-\frac{m_a^2}{2}\,\phi^2=-\frac{1}{2}A^2 m_a^2 \cos 2m_a t \ .
\label{pressure}
\end{eqnarray} 
The time scale of the oscillation is given by $m_a^{-1}$ and the cosmological time scale is $H_0^{-1}$. Since we consider $10^{-10}$ eV $\leq$  $m_a$ $\leq$ 1 keV and $H_0\sim 10^{-33}$ eV, the time scale of the oscillation is much shorter than the cosmological time scale. Hence, 
the pressure in Eq.~(\ref{pressure}) effectively vanishes on average.
Thus, the axion is known to be a good dark matter candidate.

If the occupation number becomes at the order of one, ${\cal O}(1)$, quantum coherence becomes manifest. Let us see the case of the light axion dark matter $m_a\sim 10^{-6}\,{\rm eV}$. The occupation number of the axion is estimated as
\begin{eqnarray}
\frac{\Delta N}{\Delta x^3 \Delta p^3} \sim  \frac{n}{k^3} 
= \frac{\rho_{\rm DM}}{m_a k^3} 
\sim 2\times 10^{28} \left(\frac{\rho_{\rm DM}}{0.3\,{\rm GeV}/{\rm cm}^3}\right) 
\left(\frac{10^{-6}\,{\rm eV}}{m_a}\right)^4 
\,,
\end{eqnarray}
where $n$ is volume number density and $p= k=m_av_a$ and also the velocity $v_a=0.5\times 10^{-3}$~\cite{Gorbunov:2011} are used. In this case, the occupation number is large and we regard the axion as a classical field naively. However, for the axion with the mass about 10 eV, 
the occupation number becomes
\begin{eqnarray}
\frac{\Delta N}{\Delta x^3 \Delta p^3} 
\sim {\cal O}(1) \left(\frac{\rho_{\rm DM}}{0.3\,{\rm GeV}/{\rm cm}^3}\right) 
\left(\frac{10\,{\rm eV}}{m_a}\right)^4 
\,,
\end{eqnarray}
We see the quantum coherence will be manifest. We note that occupation number is one way to distinguish the quantum object from classical object. Thus we consider the possibility to find the quantum coherence in a broad mass range from 
$10^{-10}$ eV up to $1$ keV in the following.

\section{Quantum Coherence}
\label{section3}

One of the most striking difference between classical and quantum physics is quantum coherence which is rooted in the superposition principle. That is, a single quantum state simultaneously consists of multiple states. Let us consider a two-dimensional Hilbert space spanned by a basis $\{|1\rangle$, $|2\rangle\}$. Assuming that the density operator can be written by $\rho =\left( |1\rangle\langle1| + |2\rangle\langle2| + c|1\rangle\langle2| + c^* |2\rangle\langle1|\right)/2$ where the complex number $c$ satisfies $|c|\leq1$. Hence, $c$ represents the off-diagonal elements of $\rho$. For one extreme $|c| = 1$, the state is in a pure equal superposition between the two states $|1\rangle$ and $|2\rangle$. In the other extreme, $c = 0$, the state is maximally mixed and in no superposition. This tells us that the off-diagonal element $c$ describes the superposition, and it seems reasonable to take the quantity $|c|$ as a measure of the degree of superposed states between the two states $|0\rangle$ and $|1\rangle$. This observation generalizes to pairs of orthogonal subspaces, in the sense that the off-diagonal block carries the information of the superposition.

The amount of coherence present in a system can be
quantified with a coherence measure $C(\rho)$ 
 such that
\begin{eqnarray}
C(\rho) \geq 0 \ , 
\end{eqnarray}
with equality if and only if $\rho$ is incoherent. A simple
example of such a function is given by the $\ell_1$-norm of
coherence, which is equal to the sum of the modulus
of the  non-diagonal elements
\begin{eqnarray}
C(\rho) = \sum_{i\neq j} \big| \rho_{ij} \big|\,,
\label{measure}
\end{eqnarray}
with values ranging between 0 for an incoherent state
and $d-1$ for the maximally coherent $d$-dimensional pure
state
\begin{eqnarray}
\big| \psi \big>  = \frac{1}{\sqrt{d}}\sum_{i=0}^{d-1} \big| i \big> \ .
\end{eqnarray}
For the above example of two states $|1\rangle$ and $|2\rangle$, the coherence measure is   $C(\rho)=|c|$.
In order to harvest quantum coherence from a coherent system to an incoherent system, it is necessary to make them interact through a completely positive and trace preserving quantum operation as we will see next. 

\section{Harvesting coherence of axion dark matter}
\label{section4}

\subsection{The inertial Unruh-DeWitt detector}

Let us consider an observer carrying a point-like Unruh-DeWitt detector~\cite{Unruh:1976db,DeWitt}. The Unruh-DeWitt detector consists of a two-level of quantum system with the energy gap $\Omega$, interacting locally with axions along the detector's worldline. 
In this paper, we only consider a constant velocity observer. It is straightforward  to extend the analysis to accelerating observer.
The Hamiltonian of the detector is given by 
\begin{eqnarray}
\hat{H}_{\rm D} = \frac{\Omega}{2} \left(  \big| e \rangle \langle e \big| -   \big| g \rangle \langle g \big| \right)\,,
\end{eqnarray}
where   $\big| g \rangle$ and  $\big| e \rangle$ are ground and excited states, respectively.
The worldline ${\bm x}(\tau)$ of the detector is parametrized by proper time $\tau$. 
The interaction of Hamiltonian of the detector and the axion field is conveniently describe in terms of the detector's proper time given by
\begin{eqnarray}
\hat{H}_{\rm I} (\tau) = \lambda \chi(\tau)  \hat{\mu} (  \tau )  \otimes \partial_\tau \hat{\phi} (\tau)  \  ,
\label{interaction}
\end{eqnarray}
where $\lambda $ is a coupling constant and the monopole moment operator $\hat{\mu}$ and the switching function $\chi$ of the detector are respectively given by 
\begin{eqnarray}
&&\hat{\mu} ( \tau ) = e^{i\Omega \tau}  \big| e \rangle \langle g \big| + e^{- i\Omega \tau}  \big| g \rangle \langle e \big| \,,\label{monopole}\\
&&\chi (\tau )  = \frac{1}{\pi T}\exp\left[ - \frac{\tau^2}{\pi T^2}  \right]\,.
\label{switching}
\end{eqnarray}
Note that the switching function describes the way the interaction of the duration $T$ is switched on and off.
The axion has the symmetry under the transformation $\phi\rightarrow\phi$ + constant, which requires the derivative coupling in the interaction Hamiltonian~(\ref{interaction}). 

The axion field is quantized as
\begin{eqnarray}
\hat{\phi}  (x)  = \int \frac{d^3{\bf k}}{\sqrt{(2\pi)^3 2\omega_k}} 
\left[  \hat{a}_{\bf k} \,e^{ -i\omega_k t+i{\bf k} \cdot {\bf x}}  +  \hat{a}^\dag_{\bf k}\,e^{i\omega_k t -i{\bf k} \cdot {\bf x}} \right]\,,
\end{eqnarray}
with creation and annhilation operators $ a_{\bf k}, a^\dagger_{\bf k}$. Note that four-vectors and three-vectors are denoted by bold-italic and boldface type, respectively.

\subsection{Harvesting quantum coherence }

Let us suppose that the initial state $\hat{\rho}^\text{in} $ before the interaction between the detector and the axion is switched on is given in a separable state
\begin{eqnarray}
\hat{\rho}^\text{in}=\big| g \rangle \langle g \big|  \otimes \hat{\sigma}_\phi   \  ,
\end{eqnarray}
where $\hat{\sigma}_\phi $ is an initial state of the axion. In the interaction picture, the time evolution of the system is described by the unitary operator
\begin{eqnarray}
\hat{U} &=& {\rm T} \exp\left[   -i \int \hat{H}_{\rm I} \,d\tau    \right] \,,   \nonumber \\
&=& 
\text{T} \exp\left[ -i  \lambda   \big| e \rangle\langle g \big|  \otimes  \hat{\Phi}^\dagger  
    -   i  \lambda   \big| g \rangle \langle e \big|  \otimes  \hat{\Phi} \right] \,.
\end{eqnarray}
Here we used Eqs~(\ref{interaction}) and (\ref{monopole}) and defined
\begin{eqnarray}
\hat{\Phi} \equiv \int^\infty_{-\infty} \chi(\tau)  e^{-i \Omega   \tau }   \partial_\tau \hat{\phi}_f (x(\tau)) d\tau \,,
\label{Phi}
\end{eqnarray}
where the smeared axion field $\hat{\phi}_f$ on the detector's center of mass worldline ${\bm x}(\tau)=(t(\tau),{\bf x}(\tau))$ is defined by
\begin{eqnarray}
\hat{\phi}_f({\bm x}(\tau))&\equiv&\int f({\bm\xi})  \hat{\phi} (x(\tau , {\bm\xi} )) d^3 {\bm\xi} \,,\nonumber\\
 &=&\int \frac{d^3 {\bf k}}{\sqrt{(2\pi)^3 2\omega_k}}
 \left[  F(\tau,{\bf k})\hat{a}_{\bf k} \,e^{ -i\omega_k t+i{\bf k} \cdot {\bf x}}  +  F^*(\tau,{\bf k})\hat{a}^\dag_{\bf k}\,e^{i\omega_k t -i{\bf k} \cdot {\bf x}} \right]\,,
 \label{phif}
\end{eqnarray}
with
$
{\bf x} (\tau , {\bm \xi} )  = {\bf x}(\tau) + {\bm\xi}\,,
$
and
\begin{eqnarray}
 F(\tau, {\bf k})\equiv\int d^3 {\bm\xi} \,f ( {\bm\xi} )  e^{ i {\bf k} \cdot  {\bm\xi} }   \,.
\end{eqnarray}
Here, we introduced a spatial profile or smearing function $f ( {\bm\xi} ) $ supported on a finite spatial region, in order to take into account the corrections coming from the finite size of the detector. The smearing function controls the interaction between the detector at each point in the region and the axion field~\cite{Schlicht:2003iy, Louko:2006zv,Pozas-Kerstjens:2016rsh}.
The smearing function $f$ is chosen as
\begin{eqnarray}
f ({\bm \xi })  = \frac{1}{(\pi R)^3}\exp\left[ - \frac{{\bm \xi }^2}{\pi R^2}  \right] \  ,
\label{smearing}
\end{eqnarray}
where $R$ represents the size of the detector.  The solution of the Liouville-von-Neumann equation for the coupled system $\partial_\tau\hat{\rho}(\tau)=-i[H_{\rm I},\hat{\rho}(\tau)]$ is found to be
\begin{eqnarray}
\hat{\rho}^\text{f}&=&\hat{U} \hat{\rho}^\text{in}\hat{U}^\dag \,\\
&\sim&\left(
    \begin{array}{ccc}
      \hat{\sigma}_\phi- \frac{\lambda^2}{2}(\hat{\sigma}_\phi\text{T}[\hat{\Phi}\hat{\Phi}^\dagger]+(\text{T}[\hat{\Phi}\hat{\Phi}^\dagger])^\dagger \hat{\sigma}_\phi)
      &   i \lambda\hat{\sigma}_\phi \hat{\Phi} \\
      -i \lambda \hat{\Phi}^\dagger \hat{\sigma}_\phi
       &  \lambda^2\hat{\Phi}^\dagger \hat{\sigma}_\phi \hat{\Phi}
    \end{array}
  \right)\, , 
  \label{perturbation}
\end{eqnarray}
where the superscript f of $\hat{\rho}$ denotes the final state and the higher order terms were neglected.
By tracing out the axion degrees of freedom, the reduced density matrix of the detector reads
\begin{eqnarray}
  \hat{\rho}^\text{f}_{\rm D} = \left(
    \begin{array}{ccc}
      1- \lambda^2 {\rm tr}(\hat{\Phi}^\dagger \hat{\sigma}_\phi \hat{\Phi} ) 
      &   i \lambda {\rm tr}(\hat{\Phi} \hat{\sigma}_\phi  )  \\
      -i \lambda {\rm tr}(\hat{\Phi}^\dagger \hat{\sigma}_\phi  )  
       &  \lambda^2 {\rm tr}(\hat{\Phi}^\dagger \hat{\sigma}_\phi \hat{\Phi} )  
    \end{array}
  \right)\,,
\end{eqnarray}
where ${\rm tr}\sigma_\phi=1$ is used. Note that the off-diagonal elements of ${\hat\rho}^\text{f}_{\rm D}$ appear due to the coupling $\lambda$.

The coherence measure Eq.~(\ref{measure}) is obtained by summing the modulus of the off-diagonal elements of $\hat{\rho}^\text{f}_{\rm D}$, which leads to
\begin{eqnarray}
C= 2 \lambda  \big| {\rm tr} \left( \hat{\Phi}\hat{\sigma}_\phi \right)\big|   \  ,
\label{measure2}
\end{eqnarray}
where $\hat{\Phi}$ in Eqs.~(\ref{Phi}) and (\ref{phif}) is calculated as
\begin{eqnarray}
\hat{\Phi}=\int \frac{d^3 {\bf k}}{\sqrt{(2\pi)^3 2\omega_k}}\left(
F_-({\bf k})\hat{a}_{\bf k} + F^*_+({\bf k})\hat{a}^\dag_{\bf k}\right)\,,
\label{Phi2}
\end{eqnarray}
with the quantity
\begin{eqnarray}
F_{\pm} ({\bf k})\equiv\int^\infty_{-\infty} \chi(\tau)  e^{\pm i \Omega   \tau }   
\partial_\tau \left( F (\tau, {\bf k}) e^{ i {\bm k} \cdot  {\bm x}(\tau) }  \right)    d\tau \,.
\end{eqnarray}
Plugging the above Eq.~(\ref{Phi2}) into Eq.~(\ref{measure2}), the coherence measure is found to be
\begin{eqnarray}
C= 2 \lambda  \Big| \int \frac{d^3 {\bf k}}{\sqrt{(2\pi)^3 2\omega_k}} 
\left[  F_{-} ({\bf k}) a({\bf k})    + F_{+}^*  ({\bf k}) a^* ({\bf k})    \right] \Big|   \,,
\label{measure3}
\end{eqnarray}
where we used axion coherent states $\hat{a}_{\bf k}|\phi\rangle=a({\bf k})|\phi\rangle\,.$
Indeed, the expectation value of the annihilation operator $\hat{a}_{\bf k}$ is given by
\begin{eqnarray}
a({\bf k}) =\langle\phi|\hat{a}_{\bf k}|\phi\rangle= {\rm tr} \left(\hat{a}_{\bf k} \hat{\sigma}_\phi \right)   \  .
\end{eqnarray}
The expectation value of the quantized axion field with the complex amplitude is expected to become
\begin{eqnarray}
{\rm tr} ( \hat{\phi}(x)\hat{\sigma}_\phi ) &=& 
\int \frac{d^3 {\bf k}}{\sqrt{(2\pi)^3 2\omega_k}} 
\left[  {\rm tr}(\hat{a}_{\bf k}\hat{\sigma}_\phi) \,e^{ -i\omega_k t+i{\bf k} \cdot {\bf x}}  +  {\rm tr}(\hat{a}^\dag_{\bf k}\hat{\sigma}_\phi)\,e^{i\omega_k t -i{\bf k} \cdot {\bf x}} \right]\,,\nonumber\\
&=& A \cos ( \omega_p t - {\bf p} \cdot {\bf x} -\theta )     \,.
\label{axion2}
\end{eqnarray} 
Hence, we take the complex amplitude of $a({\bf k})$ of the form:
\begin{eqnarray}
a({\bf k}) =   \sqrt{\frac{(2\pi)^3 \omega_p}{2} } \, a_0\,\delta^{3}({\bf k} -{\bf p}) \  
\label{eq:ak},
\end{eqnarray}
where $\omega_p = \sqrt{{\bf p}^2 +m_a^2}$ with the momentum $\bf p$ of the axion, and $a_0=Ae^{i\theta}$ with the amplitude of the axion $A$ and an initial phase $\theta$.
Since the mass of axion is much larger than its momentum $m_a\gg{\bf p}$, its energy satisfies $\omega_p\approx m_a$. Then we see Eq.~(\ref{axion2}) agrees with Eq.~(\ref{axion}).
Substituting the complex amplitude Eq.~\eqref{eq:ak} into the coherence measure Eq.~(\ref{measure3}), we find 
\begin{eqnarray}
C= \lambda  \Big| F_{-} ({\bf p}) a_0   + F_{+}^*  ({\bf p}) a^*_0  \Big|   \  \label{eq:C}. 
\end{eqnarray}

Now, let us consider whether an inertial detector can harvest coherence from the axion field. The detector is moving along a worldline with a constant four-velocity ${\bm u}$ such as
\begin{eqnarray}
{\bm x} (\tau) = {\bm u} \tau, \ \label{eq:x} 
\end{eqnarray}
where ${\bm u}= \gamma (1, {\bf v}) $ with the Lorentz factor $\gamma=1/\sqrt{1-{\bf v}^2}$. 
We assume that the size of detector is much smaller than the de Broile wavelength of the axion. 
Then the detector behaves like a point particle and the smearing function $ f({\bm\xi})$ in Eq.~(\ref{smearing}) is approximated by the delta function  $ \delta^3 ({\bm\xi}) $ in the limit $R \rightarrow 0$. 
The function $F_\pm ({\bf p})$ is computed as 
\begin{eqnarray}
F_\pm ({\bf p}) 
&=& 
\int^\infty_{-\infty} \chi(\tau)  e^{\pm i \Omega   \tau }  \partial_\tau \left( F ({\bf p} , \tau) e^{ i {\bm p} \cdot  {\bm x}(\tau) }  \right)    d\tau\,,  \nonumber \\
&\sim &  
\int^\infty_{-\infty} \chi(\tau)  e^{\pm i \Omega   \tau }  \partial_\tau \left( e^{ i  {\bm p} \cdot  {\bm x}(\tau) }  \right)    d\tau \,, \nonumber \\
&=& 
i {\bm p} \cdot {\bm u} \,  e^{ - \frac{\pi}{4} ({\bm p} \cdot {\bm u} \pm \Omega)^2 T^2 }\,,
\end{eqnarray}
where $F({\bf p},\tau) \sim \int d^3{\bm\xi}\delta^3(\bm\xi)e^{i{\bm p}\cdot{\bm\xi}}\sim 1$ is used in the second line and Eqs.~(\ref{switching}) and \eqref{eq:x} are used in the third line.  
Plugging the above back into Eq.~(\ref{eq:C}), we find that the coherence measure becomes finite such as
\begin{eqnarray}
C=  \lambda  A | {\bm p} \cdot {\bm u}| e^{-\frac{\pi}{4} ( ({\bm p}\cdot {\bm u})^2 + \Omega^2 ) T^2 }  \Big| e^{i\theta + \frac{\pi}{2} {\bm p}\cdot {\bm u}\,\Omega T^2} +  e^{-i\theta - \frac{\pi}{2} {\bm p}\cdot {\bm u}\,\Omega T^2}  \Big|   \  \label{eq:C1}. 
\end{eqnarray}
Since $C=0$ for $\lambda=0$, the result tells us that the detector can harvest the coherence from the axion field through the interaction.

Next we quantify how much the detector can harvest the coherence from the axion field. The coherence measure becomes the maximum $C_\text{max}$ and minimum $C_\text{min}$ for the initial phase $\theta=0, \pi/2$, respectively, which are expressed as 
\begin{eqnarray}
C_\text{max}&=&2\lambda  A | {\bm p} \cdot {\bm u}| e^{-\frac{\pi}{4} ( ({\bm p}\cdot {\bm u})^2 + \Omega^2 ) T^2 }  \cosh \bigl[ \frac{\pi}{2} {\bm p}\cdot {\bm u}\,\Omega T^2 \bigr],
\\ 
C_\text{min}&=&2\lambda  A | {\bm p} \cdot {\bm u}| e^{-\frac{\pi}{4} ( ({\bm p}\cdot {\bm u})^2 + \Omega^2 ) T^2 }  \, \Bigl| \sinh \bigl[ \frac{\pi}{2} {\bm p}\cdot {\bm u}\,\Omega T^2 \bigr] \Bigr|.
\end{eqnarray}
If the interaction duration becomes long enough ($T\rightarrow\infty$), the coherence measure decays exponentially,
\begin{eqnarray}
C=\lambda  A | {\bm p} \cdot {\bm u}|e^{-\frac{\pi}{4}({\bm p} \cdot {\bm u}-\Omega)^2T^2}\,.
\end{eqnarray}
For the axion, ${\bm p}\sim m_a$, we have $| {\bm p} \cdot {\bm u}| \sim m_a \gamma$ and the amplitude of the axion $A$ is estimated as in Eq.~(\ref{A}), so the coherence measure is written by
\begin{eqnarray}
C=\sqrt{2\rho_{\rm DM} }\,\lambda\gamma \exp\left[  {-\frac{\pi}{4} ( m_a \gamma- \Omega )^2 T^2 }  \right]
  \ .  \label{eq:result1}
\end{eqnarray}
By using the data $\rho_{\rm DM} = 0.3\,{\rm  GeV}/{\rm cm}^3$~\cite{Gorbunov:2011} and choosing the energy gap appropriately ($\Omega\sim m_a\gamma$), we  get
\begin{eqnarray}
    C  \sim  10^{-3} \gamma \,\left(\lambda\ {\rm eV}^2 \right)
  \ .  \label{eq:result2}
\end{eqnarray}
 For a sufficiently large coupling, this would be measurable.
Hence, it might be possible to harvest quantum coherence from the axion dark matter. The mass $m_a$ we can probe is determined by the 
energy gap $\Omega$ of the detector.

\subsection{Two-level of electron system in an atom as the Unruh-DeWitt detector}

As an example, we consider a two-level of electron system in an atom as the Unruh-DeWitt detector. The interacting action between the axion and the electron reads
\begin{eqnarray}
S = \frac{1}{2}g_{ae} \bar{\psi}_e \gamma^\mu\gamma_5 \psi_e  \partial_\mu \phi \ ,
\end{eqnarray}
where $\psi_e$ is the electron field and $g_{ae}$ is a coupling constant. $\gamma^\mu$ is $4\times4$ gamma matrices and $\gamma_5=i\gamma^0\gamma^1\gamma^2\gamma^3$.
It is known that the axion coupling to an electron is 
constrained to be less than $g_{ae} = 3.3 \times 10^{-13} m_e^{-1}$ from axion bremsstrahlung in globular cluster red giants ~\cite{Marsh:2015xka}.  If we naively translate it to the present case, 
$\lambda = 3.3\times 10^{-13} m_e^{-1} T$, where the electron mass $m_e =0.5$ MeV,
\begin{eqnarray}
C=2.2\times 10^{-6}\gamma\, \left(T/1{\rm s}\right) \ ,
\end{eqnarray}
where $T$ is the observation time.
For $T=10$ s, $C$ is of the order of $2.2\times 10^{-5}\gamma$ and a bit small to detect. However, if we consider an efficient detector such as a magnon detector~\cite{Ikeda:2021mlv}, the large number of atoms in a condensed matter system can increase the sensitivity because 
the coherent interaction could enhance the strength of the coupling constant. Thus, there is a chance to detect the quantum coherence of the axion and then probe the axion dark matter.

\
\section{Conclusion}
\label{section5}

We have quantified the coherence measure of a detector in order to see if the detector can harvest the quantum coherence from the axion dark matter. We employed the Unruh-DeWitt detector which interact locally with the axion dark matter. We calculated $\ell_1$ norm as a coherent measure of the Unruh-DeWitt detector. We found that the Unruh-DeWitt detector can harvest the coherence in Eq.~(\ref{eq:result2}):
$$C  \sim  10^{-3} \gamma \,\left( \lambda\ {\rm eV}^2 \right) \,,$$
from the axion dark matter. Since the coherence measure in Eq.~(\ref{eq:result1}) has a peak at
$\Omega\sim m_a\gamma$, the axion mass $m_a$ we can probe is determined by the energy gap $\Omega$ of the detector.   
Our perturbation method to obtain the reduced density matrix of the detector in (\ref{perturbation}) is still applicable even for the coupling $\lambda\sim{\cal O}(1)\, {\rm eV}^{-2}$. 
We note that the $\lambda$ is a dimensionful quantity, so the expansion parameter in our method should be a dimensionless quantity $\lambda \text{tr}(\hat{\Phi} \hat{\sigma}_\phi)$.
Thus, the small coherence measure \eqref{measure2}
$C\sim \lambda \text{tr}(\hat{\Phi} \hat{\sigma}_\phi)\sim10^{-3}$ for the coupling guarantees the validity of our analysis. For such a coupling, our approach may provide a new method of probing the axion dark matter. 

As a concrete example, we considered a two-level of electron system in an atom as the  detector. 
In this case, we obtained the coherence measure $C=2.2\times 10^{-6}\gamma\, \left(T/1{\rm s}\right)$. 
We need to consider more detailed setup for the detectors. 
In particular, we should study more efficient  detectors that enhance the sensitivity by using the condensed matter. 
We leave this issue for future work.

\section*{Acknowledgments}
S.\,K. was supported by the Japan Society for the Promotion of Science (JSPS) KAKENHI Grant Number JP18H05862.
J.\,S. was in part supported by JSPS KAKENHI Grant Numbers JP17H02894, JP17K18778, JP20H01902. This research was supported by the Munich Institute for Astro- and Particle Physics (MIAPP) which is funded by the Deutsche Forschungsgemeinschaft 
(DFG, German Research Foundation) under Germany's Excellence Strategy
-EXC-2094-390783311.


\begin{thebibliography}{99}
\bibitem{Stott:2020gjj}
M.~J.~Stott,
%``Ultralight Bosonic Field Mass Bounds from Astrophysical Black Hole Spin,''
[arXiv:2009.07206 [hep-ph]].

%\cite{Aoki:2016kwl}Full journal details should be provided for references [14], [23], [25],
[29], [32], [34] which have been peer-reviewed.Full journal details should be provided for references [14], [23], [25],
[29], [32], [34] which have been peer-reviewed.
\end{thebibliography}
\end{document}